# BEAM DYNAMICS AND WAKE-FIELD SIMULATIONS FOR THE CLIC MAIN LINACS

V. F. Khan, R.M. Jones, Cockcroft Institute, Daresbury, WA4 4AD, UK
and University of Manchester, Manchester, M13 9PL, UK

*Abstract*

The CLIC linear collider aims at accelerating multiple bunches of electrons and positrons and colliding them at a centre of mass energy of 3 TeV. These bunches will be accelerated through X-band linacs, operating at an accelerating frequency of 12 GHz. Each beam readily excites wake-fields within the accelerating cavities of each linac. The transverse components of the wake-fields, if left unchecked, can dilute the beam emittance. The present CLIC design relies on heavy damping of these wake-fields in order to ameliorate the effects of the wake-fields on the beam emittance. Here we present initial results on simulations of the long-range wake-fields in these structures and on beam dynamics simulations. In particular, detailed simulations are performed, on emittance dilution due to beams initially injected with realistic offsets from the electrical centre of the cavities.

## INTRODUCTION

The CLIC scheme aims at colliding electrons and positrons at a centre of mass energy of 3 TeV. The main accelerating cavities of CLIC are normal conducting copper structures, designed to operate at 12 GHz. As the beam transits these accelerating cavities it receives a tranverse momentum kick which has the potential to result in serious beam disruption. This kick is proportional to the $a^3$, where "a" is the iris dimension. It is interesting to note that for the CLIC scenario the average iris is ~3.0 mm compared to 35 mm for the ILC superconducting cavities (which operate at an L-band frequency of 1.3 GHz). Thus, the kick imparted to the beam in the CLIC design is a factor of 1600 greater than that of the ILC cavities. Clearly, the wake-field will require careful suppression and the impact on emittance dilution will necessitate a beam dynamics study including realistic fabrication tolerances.

The present baseline design for CLIC relies on heavy damping (with Qs as low as 10) in order to suppress these wake-fields. The wake-field suppression in this case entails locating the damping materials in relatively close proximity to the location of the accelerated beam. Here a strategy originally employed for the NLC [1] in which moderate damping is imposed (Q~500) together with detuning of the characteristic modes of the structure is used. The potential advantage of this alternative method lies in the ability to locate the damping materials outside the immediate vicinity of the beam and to provide a means of diagnosing the location of the beam and cell misalignments from the radiation at the damping ports [2].

We have already considered a design in which we prescribed a Gaussian dipole mode distribution with an optimised bandwidth of 3.3 GHz for the first band together with interleaving of successive structures [3]. The detuning resulted in a wake-field which was suppressed by almost two order of magnitude at the first trailing bunch in the CLIC train of 312 bunches. In this work we focus on the parameters of the present baseline design known as CLIC_G which consists of 24 cells. We retain the dimensions of the end cells and the intermediate cells are modified with a view to enforcing a Gaussian distribution in the kick factor weighted distribution [1]. Enforcing a fixed geometry to the end cells results in a reduced dipole mode bandwidth. As a consequence of this fixed dipole bandwidth the wake-field at the first few trailing bunches is insufficiently damped. Interleaving successive structures results in an improved overall suppression in the wake-field. Nonetheless, even with 8-fold interleaving of structures the envelope of the wake-field is still unsatisfactory. To ameliorate the effect of this wake-field on emittance dilution we have modified the structure geometry with a view to locating the first few trailing bunches at the zero crossing point in the wake-field. To assess the practicality of this method we have undertaken a series of beam dynamics simulations entailing tracking the beam through the complete CLIC main linac. We present initial results on this study herein. The RMS of the sum wake-field also provides evidence as to whether beam break up (BBU) is occurring and we study this parameter also.

## BASELINE CLIC STRUCTURE

The fundamental parameters of the present baseline structure known as CLIC_G are delineated in Table 1. From an analysis of electrical breakdown the group velocity of the fundamental mode should be kept as small as is commensurate with a practical filling time. With this in mind, the geometry of the end cells is invariant to within a factor of ~10%. Intermediate cells are varied in order to enforce a Gaussian distribution of the kick factor weighted distribution. The resulting wake-field for a structure of N cells is calculated from the modal summation:

$$W(t) = 2\,\mathrm{Im}\left\{\sum_{p=1}^{N} K_p \exp\left[i\omega_p t\left(1+\frac{i}{2Q_p}\right)\right]\right\} \quad (1)$$

where for the $p^{th}$ mode $\omega_p/2\pi$, $K_p$ is the coupled mode frequency and kick factor, respectively. A damping factor $Q_p$ has also been incorporated. Furthermore, it is convenient to display the maximum excursion in the wake-field and this is obtained by replacing the Im factor in Eq. (1) with the absolute value.

Table 1: Parameters of updated CLIC baseline structure CLIC_G [4].

| Structure | CLIC_G |
|---|---|
| Frequency (GHz) | 12 |
| Average iris radius/wavelength <a>/λ | 0.11 |
| Input /Output iris radii (mm) | 3.15, 2.35 |
| Input /Output iris thickness (mm) | 1.67, 1.0 |
| Group velocity (% c) | 1.66, 0.83 |
| Number of cells per cavity | 24 |
| Bunch separation (rf cycles) | 6 |
| Number of bunches in a train | 312 |

Our design focussed on a 24 cell structure and 8-fold interleaving of the structure (making the structure effectively 192 cells). We investigated a Gaussian kick factor weighted distribution with a standard deviation $\sigma$ from the central value $\omega_c/2\pi$. The optimised design features a bandwidth of $\Delta\omega/2\pi = 3\sigma = 5.8\%$ $\omega_c = 1$ GHz. The envelope of the wake-field for a single, non-interleaved structure is illustrated in Fig. 1. For this bandwidth, the wake-field is insufficiently damped as at the first trailing bunch it is 74% of the peak value (~177 V/pC/mm/m). Interleaving structures reduces the amplitude of the wake-field experienced by the first trailing bunch is reduced to ~ 26% of the peak value. The amplitude of the wake-field for the first 4 trailing bunches is displayed in Fig 2.

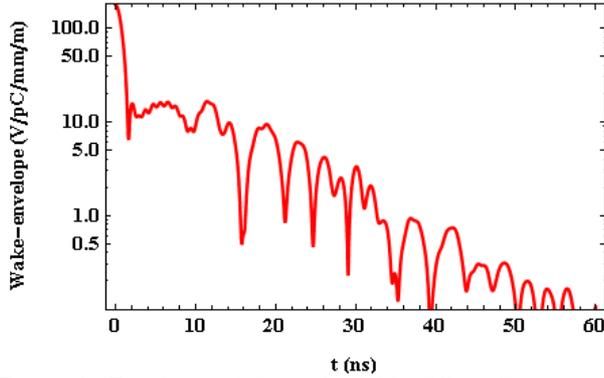

Figure 1: Envelope of the wakefield of 24 cell structure, CLIC_ZC with Q ~500.

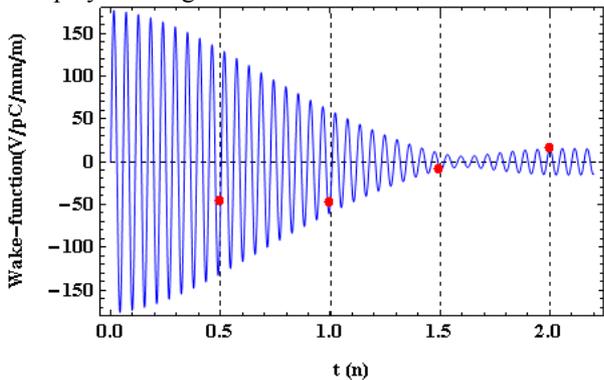

Figure 2: Wake-function of a modified CLIC_G structure, representing the wake at first four trailing bunches.

## PERTURBED BASELINE STRUCTURE

The wake felt by the bunch is obtained with Eq. 1. The oscillations in the wake-field amplitude are due to the corresponding sinusoidal term. We adjust the mode frequencies to force the bunches to be located at the zero crossing in the wake-field. This can be achieved in a straightforward manner for the first few trailing bunches by shifting the mode frequencies of all cells. In practise this is achieved by systematically shifting all cell dimensions (aperture and cavity radius). The parameters of 7 fiducial cells of a 24 cell structure, which we refer to as CLIC_ZC, are presented in Table 2. Here a is the iris radius, b is the cavity radius, t is the iris thickness, $v_g/c$ is the group velocity of the fundamental mode and $\omega_1/2\pi$ is the first band dipole frequency. The ratio of average iris radius to wavelength ($<a>/\lambda$) is 0.103. In order to maintain the group velocity of CLIC_G it was necessary to change the first cell iris radius by 160μm and the last cell iris radius by 220μm.

The group velocities of the fundamental mode are retained within the tolerable limits of 10% [5] by modifying the iris thickness of the cells accordingly. The wake-envelope of the detuned CLIC_ZC structure is relatively invariant with respect to the shifted perturbation in cell parameters.

Table 2. Parameters of the detuned CLIC_ZC structure.

| Cell. | a (mm) | B (mm) | t (mm) | Vg/c (%) | $\omega_1/2\pi$ (GHz) |
|---|---|---|---|---|---|
| 1 | 2.99 | 9.88 | 1.6 | 1.49 | 17.57 |
| 4 | 2.84 | 9.83 | 1.4 | 1.38 | 17.72 |
| 8 | 2.72 | 9.80 | 1.3 | 1.29 | 17.85 |
| 12 | 2.61 | 9.78 | 1.2 | 1.18 | 17.96 |
| 16 | 2.51 | 9.75 | 1.1 | 1.06 | 18.07 |
| 20 | 2.37 | 9.73 | 0.96 | 0.98 | 18.2 |
| 24 | 2.13 | 9.68 | 0.7 | 0.83 | 18.40 |

However, there is a considerable difference in the wake-field experienced by the first trailing bunch, which is 1.7 % of peak value in this case. The corresponding wake-field is illustrated in the Fig. 3 for the first 4 trailing bunches.

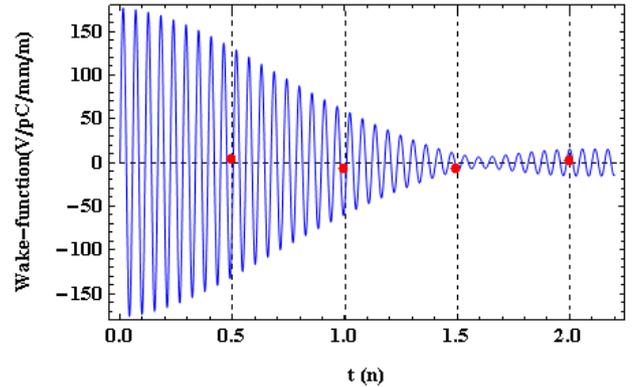

Figure 3: Wake-field of the CLIC_ZC structure. The wake-field at the location of the bunch is indicated by the dots.

In order to further improve wake-field suppression we incorporated 8-fold interleaving of the mode frequencies of successive structures and this is illustrated in Fig. 4. for a Q of 500, it is notable that the wake-field is below unity after approximately 15 ns, i.e. 30 trailing bunches.

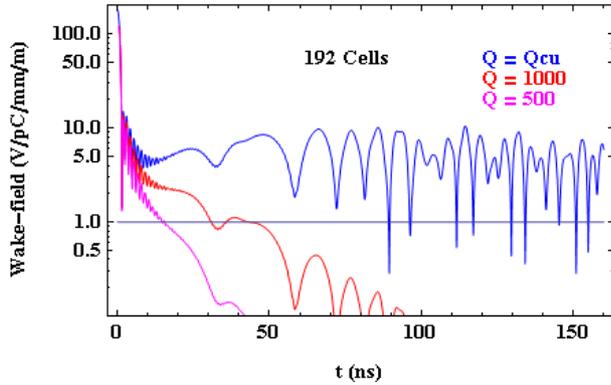

Figure 4: Envelope of wake-field for an 8-fold interleaving of dipole mode frequencies in the CLIC_ZC structure.

## BEAM DYNAMICS SIMULATIONS

The emittance dilution due to these long-range wake-fields is evaluated by tracking the progress of the beam throughout the complete linac. For this purpose we utilised the code PLACET [6]. However, an indication as to whether BBU occurs is also provided by the RMS of the sum wake-field, $S_{RMS}$ [7]. This is a direct and straightforward calculation. As a prelude to full beam dynamics tracking simulations we investigated the sensitivity of $S_{RMS}$ to small fractional errors in the bunch spacing. This is illustrated in Fig. 5 for both a single structure and an interleaved structure. At the nominal bunch spacing $S_{RMS}$ is ~33 V/pC/mm/m for a single structure and it is reduced to ~7 V/pC/mm/m for an 8-fold interleaved structure.

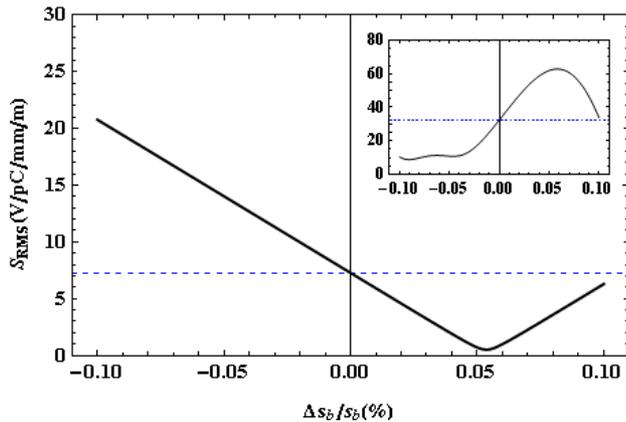

Figure 5: $S_{RMS}$ of a 24 cell structure interleaved 8-fold with a damping Q of 500. The 24 cell non-interleaved structure result is shown inset.

We also tracked the beam through the complete linac and computed the projected emittance of the beam. The results of this initial study indicate that a single, non-interleaved structure gives rise to a huge emittance growth, as expected. However, for an 8-fold interleaved structure the emittance dilution is reduced to ~ 70% at the nominal bunch spacing. This is still not acceptable from the perspective of the impact on beam luminosity. However, further optimisation may lead to a further reduction in the beam emittance. Detailed results on this study are displayed in Fig. 6. in which the emittance dilution down the linac is evaluated for several injection offsets in the beam from the centre of the linac.

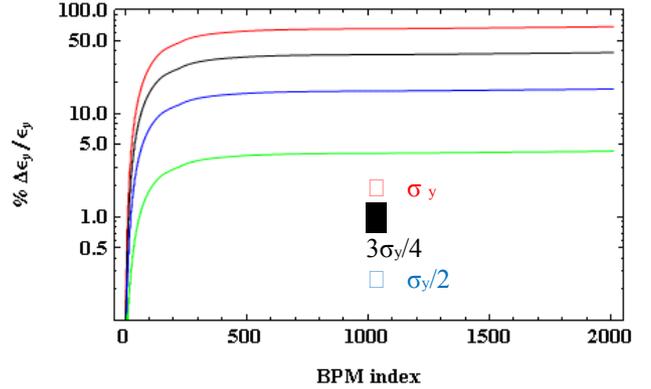

Figure 6: Emittance dilution in an 8-fold interleaved CLIC_ZC structure for various beam offsets.

## DISCUSSION

Interleaving the dipole frequencies of successive structures reduces the overall emittance dilution significantly. A beam subjected to a $\sigma_y$ injection offset results in an emittance dilution of ~70% at the nominal bunch spacing. Additional simulations conducted on the sensitivity of the emittance dilution to systematic fabrication errors indicate a strong sensitivity. However, a more careful randomisation of the zero crossing point in groups of structures has the potential to reduce the emittance dilution.

## ACKNOWLEDGMENTS

We have benefited from valuable discussions with W. Wuensch and A. Grudiev regarding the recent structures and with D. Schulte, B. Dalena and A. Latina on the beam dynamics code PLACET.